\documentclass[aps,pra,reprint,amsmath,amssymb,floatfix,superscriptaddress]{revtex4-1}

\usepackage{physics}
\usepackage{color}
\usepackage{graphicx}
\usepackage{times,psfrag}
\usepackage{dsfont}
\usepackage{dcolumn}
\usepackage{bm,bbm}
\usepackage{latexsym,amsmath,amssymb,bm,euscript}
\usepackage[normalem]{ulem}
\usepackage[hidelinks]{hyperref}
\usepackage{microtype}

\newcommand{\WSe}{WSe$_{2}$}

\newcommand{\WS}{WS$_{2}$}

\newcommand{\EF}{$E_\mathrm{F}$}
\newcommand{\hPLD}{\textit{h}-PLD}

\begin{document}
\title{%
Spin splitting and strain in epitaxial monolayer WSe$_2$ on graphene
}
\author{H. Nakamura}
\email{hnakamur@uark.edu; present address: Department of Physics, University of Arkansas, AR 72701, USA}
\affiliation{Max Planck Institute for Solid State Research, 70569 Stuttgart, Germany}
\author{A. Mohammed}
\affiliation{Max Planck Institute for Solid State Research, 70569 Stuttgart, Germany}
\author{P. Rosenzweig}
\affiliation{Max Planck Institute for Solid State Research, 70569 Stuttgart, Germany}
\author{K. Matsuda}
\affiliation{Max Planck Institute for Solid State Research, 70569 Stuttgart, Germany}
\affiliation{Nagoya University, Nagoya 464-8603, Japan}
\author{K. Nowakowski}
\affiliation{Max Planck Institute for Solid State Research, 70569 Stuttgart, Germany}
\affiliation{University of Twente, 7522 NB Enschede, Netherlands}
\author{K. K{\"u}ster}
\affiliation{Max Planck Institute for Solid State Research, 70569 Stuttgart, Germany}
\author{P. Wochner}
\affiliation{Max Planck Institute for Solid State Research, 70569 Stuttgart, Germany}
\author{S. Ibrahimkutty}
\affiliation{Max Planck Institute for Solid State Research, 70569 Stuttgart, Germany}
\author{U. Wedig}
\affiliation{Max Planck Institute for Solid State Research, 70569 Stuttgart, Germany}
\author{H. Hussain}
\affiliation{Diamond Light Source Ltd, Didcot, Oxforshire OX11 0DE, United Kingdom}
\author{J. Rawle}
\affiliation{Diamond Light Source Ltd, Didcot, Oxforshire OX11 0DE, United Kingdom}
\author{C. Nicklin}
\affiliation{Diamond Light Source Ltd, Didcot, Oxforshire OX11 0DE, United Kingdom}
\author{B. Stuhlhofer}
\affiliation{Max Planck Institute for Solid State Research, 70569 Stuttgart, Germany}
\author{G. Cristiani}
\affiliation{Max Planck Institute for Solid State Research, 70569 Stuttgart, Germany}
\author{G. Logvenov}
\affiliation{Max Planck Institute for Solid State Research, 70569 Stuttgart, Germany}
\author{H. Takagi}
\affiliation{Max Planck Institute for Solid State Research, 70569 Stuttgart, Germany}
\affiliation{Department of Physics, University of Tokyo, 113-0033 Tokyo, Japan}
\affiliation{Institute for Functional Matter and Quantum Technologies, University of Stuttgart, 70569 Stuttgart, Germany}
\author{U. Starke}
\affiliation{Max Planck Institute for Solid State Research, 70569 Stuttgart, Germany}

\begin{abstract}
We present the electronic and structural properties of monolayer \WSe\ grown by pulsed-laser deposition on monolayer graphene (MLG) on SiC.
The spin splitting in the \WSe\ valence band at $\overline{\mathrm{K}}$ was $\Delta_\mathrm{SO}=0.469\pm0.008$~eV by angle-resolved photoemission spectroscopy (ARPES).
Synchrotron-based grazing-incidence in-plane X-ray diffraction (XRD) revealed the in-plane lattice constant of monolayer \WSe\ to be $a_\mathrm{WSe_2}=3.2757\pm0.0008 \mathrm{\AA}$.
This indicates a lattice compression of $-$0.19\% from bulk \WSe. By using experimentally determined graphene lattice constant
($a_\mathrm{MLG}=2.4575\pm0.0007 \mathrm{\AA}$), we found that a 3$\times$3 unit cell of the slightly compressed \WSe\ is perfectly commensurate with a 4$\times$4 graphene lattice with a mismatch below 0.03\%, which could explain why the monolayer \WSe\ is compressed on MLG.
From XRD and first-principles calculations, however, we conclude that the observed size of strain is negligibly small to account for a discrepancy in $\Delta_\mathrm{SO}$  found between exfoliated and epitaxial monolayers in earlier ARPES.
In addition, angle-resolved, ultraviolet and X-ray photoelectron spectroscopy shed light on the band alignment between \WSe\ and MLG/SiC and indicate electron transfer from graphene to the \WSe\ monolayer. As further revealed by atomic force microscopy, the \WSe\ island size depends on the number of carbon layers on top of the SiC substrate. This suggests that the epitaxy of \WSe\ favors the weak van der Waals interactions with graphene while it is perturbed by the influence of the SiC substrate and its carbon buffer layer.
\end{abstract}

\maketitle

\section{Introduction\label{intro}}
Two-dimensional (2D) transition metal dichalcogenides (TMDs) MX$_2$ (M = Mo or W, X= S, Se, or Te) possess outstanding electronic, spin, and optical properties at thicknesses of a few layers and hold great promise for future optoelectronic and spintronic applications \cite{Mak2016,Rivera2018,Wang2018,Mak2018,Wang2012,Novoselov2016}. In the monolayer limit, the breaking of structural inversion symmetry gives rise to a large spin splitting in the top valence band located at the $\overline{\mathrm{K}}$ and $\overline{\mathrm{K}}'$ points of the surface Brillouin zone\,\cite{Xiao2012,Zhu2011,Yuan2013}. Due to time reversal symmetry, the $\overline{\mathrm{K}}$ and $\overline{\mathrm{K}}'$ valleys have opposite out-of-plane spin polarization and each valley is associated with optical selection rules of opposite chirality as well as opposite signs of Berry curvature \cite{Xiao2012}. This leads to the valley-contrasting physics of monolayer TMDs, such as optical valley polarization and the valley Hall effect \cite{Mak2016,Mak2018}. Recent advances in the application of TMDs as a quantum light source is remarkable, especially for \WSe\ \cite{Srivastava2015,He2015,Koperski2015,Chakraborty2015,Luo2018,Lu2019}, where the spin-valley degree of freedom is found to be robust also in local bound carriers \cite{Lu2019}.

\begin{figure*}[!tb]
\includegraphics[width=15cm,clip]{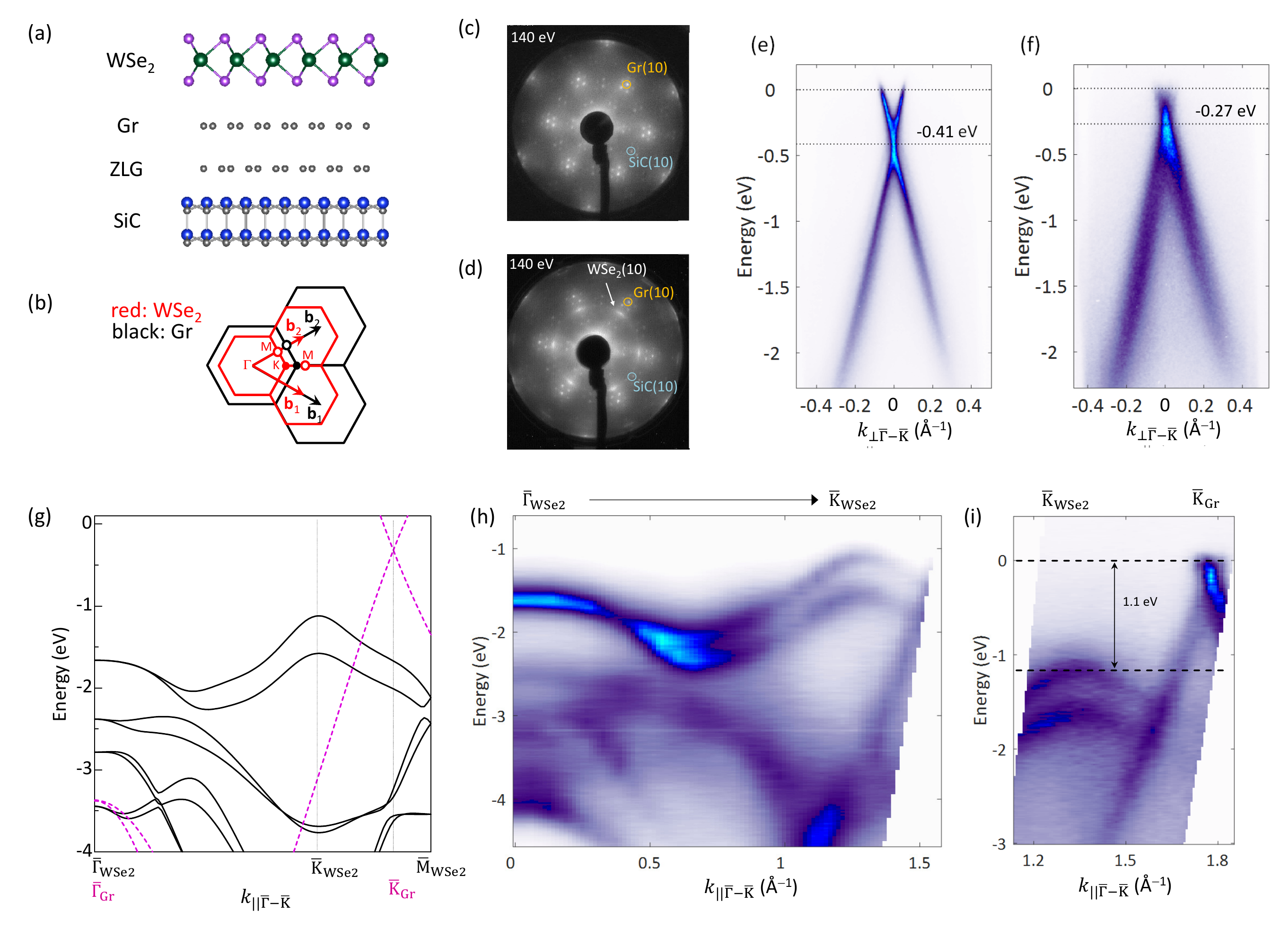}
\caption{%
ARPES of monolayer \WSe. (a) Schematic side view of the \WSe/MLG van der Waals heterostructure. (b) Reciprocal space alignment for \WSe\ and graphene. (c,d) LEED images before (c) and after (d) the growth of \WSe. (e,f) ARPES dispersion of the Dirac bands of graphene measured perpendicular to graphene's $\overline{\mathrm{\Gamma}}$-$\overline{\mathrm{K}}$ direction before (e) and after (f) the growth of \WSe. (g) Band diagram of \WSe\ (solid lines) and graphene (dashed lines) obtained from first-principles calculation. (h) ARPES data taken along the $\overline{\mathrm{\Gamma}}$-$\overline{\mathrm{K}}$ direction of \WSe\ . (i) ARPES taken at a region including the Dirac point of graphene and the top of the \WSe\ valence bands.
}
\label{arpes}
\end{figure*}

The spin splitting in the valence band at $\overline{\mathrm{K}}$ is revealed directly by angle-resolved photoemission spectroscopy (ARPES)~\cite{Zhang2013,Chiu2014,Eknapakul2014,Riley2014,Le2015,Zhang2016,Aretouli2015,
Diaz2015,Latzke2015,Riley2015,Sugawara2015,Aretouli2016,Mo2016,Aziza2017,
Forti2017,Henck2017,Agnoli2018,Wilson2017}. Spin-resolved ARPES confirmed an out-of-plane spin polarization that disappears for an even number of layers, consistent with the idea that inversion asymmetry is essential for the spin splitting~\cite{Mo2016}.
As demonstrated by theory and experiment, \WSe\ has the largest spin splitting $\Delta_\mathrm{SO}$ amongst all TMDs of 2H-type \cite{Xiao2012,Zhu2011,Le2015,Zhang2016,Wilson2017}. Le \textit{et al.}\ reported $\Delta_\mathrm{SO}=513$ meV in monolayer \WSe\ exfoliated from a bulk crystal~\cite{Le2015}, while very recent work on an exfoliated monolayer \WSe\ reported $\Delta_\mathrm{SO}=485$ meV \cite{Nguyen2019}. Zhang \textit{et al.} found $\Delta_\mathrm{SO}=475$ meV in monolayer \WSe\ grown by molecular beam epitaxy (MBE) on bilayer graphene/SiC \cite{Zhang2016}. The discrepancy in $\Delta_\mathrm{SO}$ between the MBE-grown and earlier exfoliated monolayer has been attributed to potential strain in the epitaxial TMD layer \cite{Zhang2016}. However, an evaluation of such strain in monolayer \WSe\ using a precise structural probe, such as X-ray diffraction, has thus far been missing in any of the ARPES-studied monolayer.

Besides inducing strain, the substrate beneath a TMD could have an effect on its electronic properties by affecting the growth mode or via charge redistribution at the interface \cite{Sun2017}.
TMDs on graphene represent a prototypical van der Waals (vdW) heterostructure where charge transfer could critically influence the physical properties of the TMD \cite{Froehlicher2018}.
In this regard, ARPES of the graphene $\pi$-bands before and after the creation of a vdW heterostructure could provide a direct evidence of charge transfer across the TMD/graphene interface, but no such experiment has yet been reported. Alternatively, the charge transfer can be indirectly assessed from the position of the valence band maximum of the TMD ($E_\mathrm{K}$) with respect to the Fermi level \EF.
In \WS\ grown by chemical vapor deposition on epitaxial monolayer graphene on SiC, Forti \textit{et al.}\ found $E_\mathrm{K}=-1.84$ eV~\cite{Forti2017}.
Taking into account a band gap $E_\mathrm{G}$ of 2.1 eV for pristine monolayer \WS, where \EF\ is assumed to lie mid-gap, this corresponds to a significant downshift of $\sim 0.8$ eV of the \WS\ bands. This, in turn, indicates electron transfer to \WS\ across the interface.
For MBE-grown \WSe\ on epitaxial bilayer graphene, ARPES and scanning tunneling spectroscopy (STS) yielded $E_\mathrm{K}\sim -1.1$ eV~\cite{Zhang2016}. Considering a band gap of 1.95 eV as determined by STS, this also suggests a small downshift ($\sim 100$ meV) of the \WSe\ bands, consistent with an electron transfer to the TMD layer.
On the other hand, ARPES of monolayer \WSe\ transferred to cleaved graphite yielded $E_\mathrm{K}=-0.7$ eV~\cite{Wilson2017}. Assuming the same band gap $E_\mathrm{G}$, this corresponds to \EF\ residing closer to the valence band and thus indicates a hole transfer to \WSe.
However, we note that the above results are only indirect indications of charge transfer, because the position of the Fermi level can depend on the way the respective heterostructure was prepared.
To unambiguously resolve the issue of charge transfer across the TMD/graphene interface, a comparison of ARPES measurements performed both before and after the creation of the vdW heterostructure could be highly useful.

In this paper, we clarify the electronic structure of monolayer \WSe\ grown by pulsed-laser deposition on epitaxial monolayer graphene on SiC (MLG/SiC). In particular, we address the issue of a potential strain effect on the spin splitting $\Delta_\mathrm{SO}$ by using ARPES and grazing-incidence X-ray diffraction (GIXRD) data, supported by an analysis based on first-principles calculations. The electron transfer from graphene to \WSe\ is revealed by comparing ARPES of the graphene $\pi$-bands before and after the \WSe\ deposition. Ultraviolet and X-ray photoelectron spectroscopy (UPS and XPS), which are also conducted before and after the \WSe\ growth, shed light on the band alignment between monolayer \WSe\ and graphene. Atomic force microscopy (AFM) further reveals a significant impact of the substrate morphology on the \WSe\ island size.

\section{Experiment and theory\label{exptheory}}
Monolayer graphene (MLG) on SiC was grown using the well-established recipe of sublimation growth at elevated temperatures in argon atmosphere~\cite{Emtsev2009,Forti2014}. Note that, on SiC, the graphene monolayer resides on top of a $(6\sqrt{3}\times 6\sqrt{3})\mathrm{R}30^\circ$-reconstructed carbon buffer layer (zerolayer graphene, ZLG) that is covalently bound to the SiC substrate~\cite{Riedl2010}.
\WSe\ films were grown on the thus prepared MLG/SiC substrates via hybrid-pulsed-laser deposition (\hPLD) in ultra-high vacuum (UHV)~\cite{Avaise}. This recently developed, bottom-up technique utilizes a pulsed laser to ablate transition metal targets, supported by chalcogen vapor supplied from an effusion cell, thus combining PLD and MBE. Pure tungsten (99.99 \%) was ablated using a pulsed KrF excimer laser (248\,nm) with a repetition rate of 10\,Hz, while pure selenium (99.999 \%) was evaporated from a Knudsen cell at a flux rate of around 1.5 \AA{}/s as monitored by a quartz crystal microbalance. The deposition was carried out at 450 $^{\circ}$C for three hours, followed by two-step annealing at 640 $^{\circ}$C and 400 $^{\circ}$C for one hour each. Further details on \hPLD\ can be found elsewhere~\cite{Avaise}.
GIXRD measurements were carried out at the I07 beamline of Diamond Light Source \cite{Nicklin2016}, with a photon energy of 12 keV (wavelength 1.0332 \AA) and a Pilatus 100K 2D detector (DECTRIS). The incident angle $\alpha \sim 0.2^\circ$ of the X-rays was chosen according to the critical angle of the samples, which were kept in helium atmosphere during the measurements. Topographic AFM images were acquired with a Bruker microscope in peak force tapping mode.
For photoelectron spectroscopy and LEED measurements, the freshly prepared samples were capped with a 10 nm-thick selenium layer at room temperature and transported through air into a different UHV facility, where the capping layer was removed by heating to 300 $^{\circ}$C. ARPES and UPS measurements were performed using monochromatized HeI $\alpha$ (21.22 eV) and HeII $\alpha$ (40.81 eV) photons and a 2D hemispherical analyzer equipped with a CCD Detector (SPECS Phoibos 150). The energy resolution of ARPES analyzer was 60 or 90 meV at a pass energy of 20 or 30 eV, respectively, as measured from the Fermi edge of gold at room temperature. XPS was carried out using non-monochromatized Mg K$\alpha$ (1253.6~eV) radiation. All the measurements took place at room temperature.
First-principles calculations were performed using density functional theory (DFT) as implemented in WIEN2k~\cite{wien2k} and ADF-BAND~\cite{ADF-BAND,BAND}. The generalized gradient approximation as parameterized by Perdew-Burke-Ernzerhof~\cite{PBE} was used to describe the exchange-correlation functional. The spin-orbit coupling is included in a second variational procedure (Wien2k) or in the original basis set (ADF-BAND). We used a $k$-point mesh of 16$\times$16$\times$1 and adopted a slab geometry with a 30 \AA{} gap between adjacent layers to suppress the interlayer interaction.

%\begin{figure}[!tb]
\begin{figure*}[!tb]
\includegraphics[width=14cm,clip]{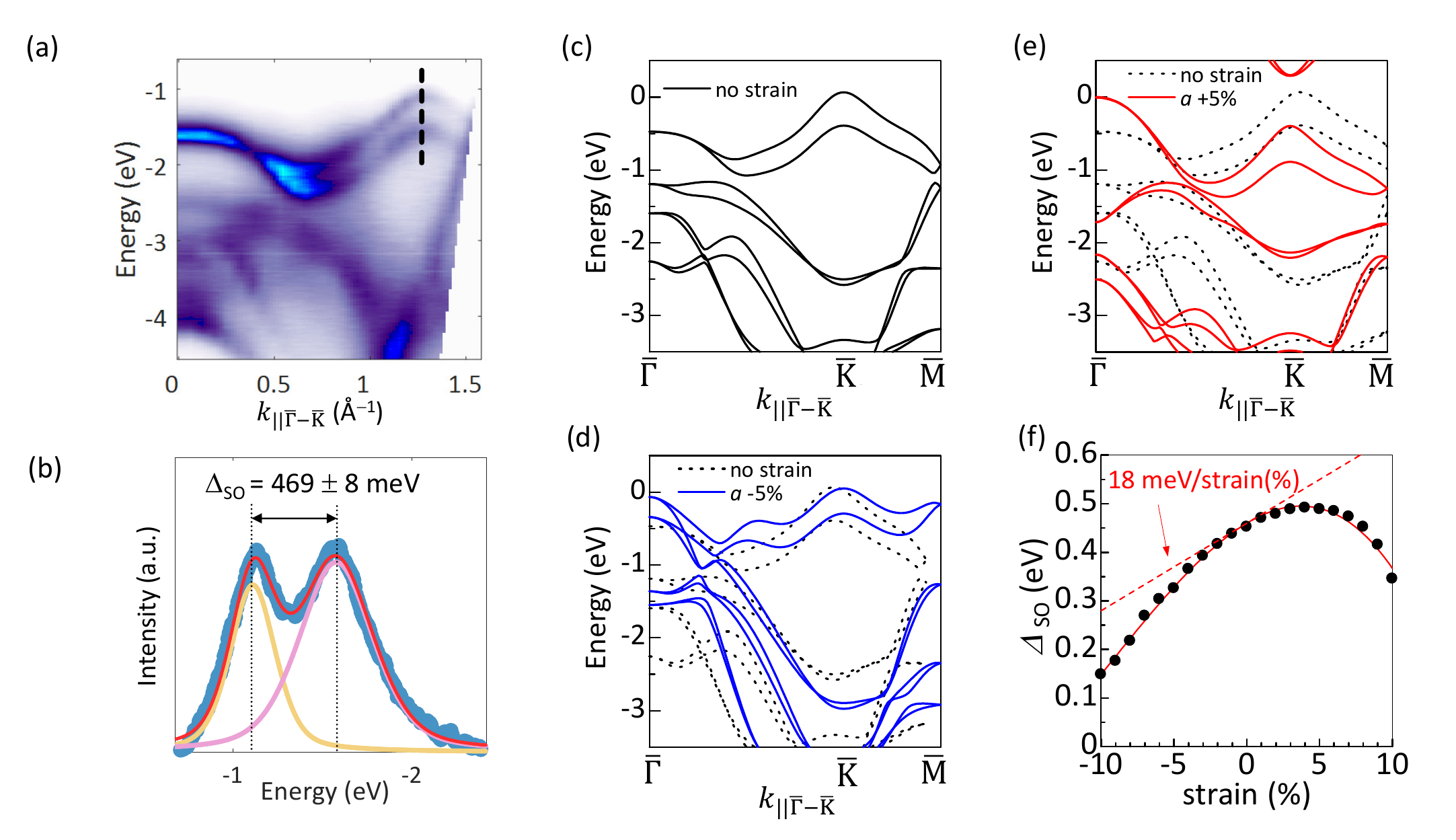}
\caption{(a) Raw ARPES data displaying the large spin splitting in the topmost \WSe\ valence band at $\overline{\mathrm{K}}$ (dashed black line). (b) EDC extracted at $\overline{\mathrm{K}}$ of \WSe\ (blue dots) together with the fit (red curve) that consists of two pseudo-Voigt functions (light orange and light purple curves). (c-e) Band structure of \WSe\ obtained from a first-principles calculation assuming (c) no strain, (d) 5 \% compressive strain (in-plane), and (e) 5 \% tensile strain (in-plane). (f) The spin splitting $\Delta_\mathrm{SO}$ as a function of in-plane strain as extracted from first-principles calculations.}
\label{fpc}
\end{figure*}

\section{Results and discussion}\label{res}
\subsection{Electronic structure and strain}\label{structstrain}

The vertical structure of the \WSe/MLG heterostack is schematically shown in Fig.~\ref{arpes}(a). Figures \ref{arpes}(c) and (d) show the LEED patterns obtained before and after the growth of monolayer \WSe\ on MLG/SiC with a coverage of approximately 50 \%, demonstrating the preferred epitaxial relationship between \WSe\ and graphene (\WSe\ [$10\overline{1}0$] $\vert \vert$ graphene [$10\overline{1}0$]).
This epitaxial relationship of the vdW heterostructure results in a reciprocal space alignment as shown in Fig.~\ref{arpes}(b). The APRES intensity map recorded along the $\overline{\Gamma\mathrm{K}}$ direction of \WSe\ and graphene is shown in Fig.~\ref{arpes}(h). The valence bands of monolayer \WSe\ are resolved with excellent quality, essentially consistent with the result of the first-principles calculation [see Fig.~\ref{arpes}(g)]. As expected from the reciprocal space alignment, the graphene $\pi$-bands with their characteristic linear dispersion in the vicinity of the Fermi level \EF\ also appear at higher parallel momenta $k_\parallel$ [see Figs.~\ref{arpes}(h) and (i)]. Note that the $\pi$-bands are shifted in energy before and after the \WSe\ deposition as revealed by the corresponding energy-momentum cuts recorded at the graphene $\overline{\mathrm{K}}$ point perpendicular to the $\overline{\Gamma\mathrm{K}}$ direction [see Figs.~\ref{arpes}(e) and (f)].
Before the \WSe\ deposition, the Dirac point is found 0.41 eV below \EF, reflecting the $n$-type doping of epitaxial graphene on SiC~\cite{Riedl2010}. After the growth of \WSe\ on top of graphene, the Dirac point has shifted to 0.27 eV below \EF. This upshift of 140 meV indicates electron transfer from graphene to the TMD monolayer which will be further discussed in Sec.~\ref{align}.

\begin{figure*}[!tb]
\includegraphics[width=17cm,clip]{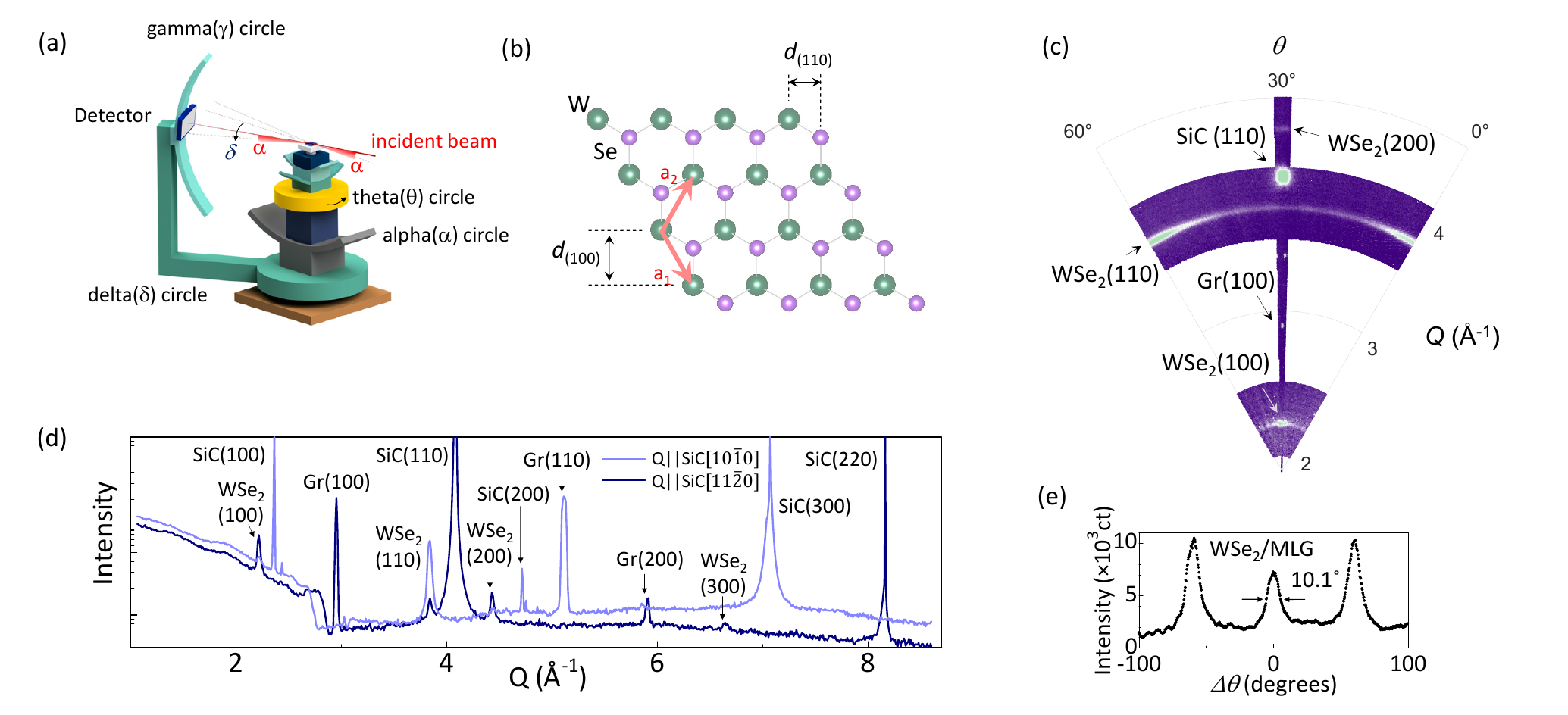}
\caption{Grazing incidence in-plane X-ray diffraction (GIXRD) of monolayer \WSe/MLG/SiC. (a) Schematic diagram of the GIXRD geometry. $\alpha \sim 0.2^\circ$ is the incidence angle of the X-ray beam. (b) In-plane structure of \WSe. The two primary in-plane lattice spacings are highlighted. (c) Reciprocal space map for \WSe\ on MLG. (d) In-plane $\delta$-$\theta$ scans along two distinct crystallographic directions ($\mathbf{Q}$ $\vert \vert$  SiC[$10\overline{1}0$] and SiC[$11\overline{2}0$]). (e) Wide-angle rocking ($\theta$) scan of the \WSe (110) Bragg reflection.}
\label{xrd}
\end{figure*}

The large spin splitting $\Delta_\mathrm{SO}$ arising in the topmost \WSe\ valence band at  $\overline{\mathrm{K}}$  due to the breaking of inversion symmetry in monolayer \WSe\ is clearly resolved in the ARPES data [see Figs.~\ref{arpes}(h), (i) and \ref{fpc}(a), (b)]. To quantify this splitting, an energy distribution curve (EDC) was extracted at the $\overline{\mathrm{K}}$ point of \WSe\ as indicated by the dashed black line in Fig.~\ref{fpc}(a). By fitting this EDC with two pseudo-Voigt curves as shown in Fig.~\ref{fpc}(b), we obtain $\Delta_\mathrm{SO}=0.469\pm0.008$ eV. The detail of the EDC analysis is shown in Supplemental Material \cite{SM}. This value is appreciably smaller than the 513 meV observed in monolayer \WSe\ exfoliated from a bulk crystal~\cite{Le2015}, while close to more recent recent result (485~meV) \cite{Nguyen2019} and MBE-grown \WSe\ (475~meV) \cite{Zhang2016}. While it is tempting to relate this difference to strain resulting from the epitaxial TMD growth, we will show in the following that the influence of strain on $\Delta_\mathrm{SO}$ is actually negligible for \WSe\ on graphene.

We first focus on the results obtained from synchrotron-based GIXRD~\cite{Avaise}. Utilizing an X-ray beam that propagates parallel to the sample surface at a critical angle of incidence $\alpha\sim0.2^{\circ}$  [see Fig.~\ref{xrd}(a)], this technique probes the in-plane structure of the \WSe\ films [see Fig.~\ref{xrd}(b)].
The in-plane reciprocal space map shown in Fig.~\ref{xrd}(c) clearly captures diffraction from monolayer \WSe. We find \WSe\ [$10\overline{1}0$] $\parallel$ graphene [$10\overline{1}0$], fully consistent with LEED [see Fig.~\ref{arpes}(d)].
The weak ring-like elongation of the \WSe\ diffraction in the reciprocal space map reflects large crystalline mosaic of monolayer \WSe\ islands with respect to rotation around the surface normal.
The wide-angle ($\pm 100^{\circ}$) rocking ($\theta$) scan for the \WSe(110) peak exhibits the expected periodicity of $60^\circ$ as shown in Fig.~\ref{xrd}(e).
To evaluate the potential strain in the epitaxial TMD film, the in-plane lattice constant $a$ of \WSe\ was extracted from the $\delta$-$\theta$ scans shown in Fig.~\ref{xrd}(d).
We find $a=3.2757\pm0.0008 \mathrm{\AA}$, which indicates a small compression of $-\,0.19$ \% with respect to the bulk reference value ($a=3.282\pm0.001 \mathrm{\AA}$~\cite{Schutte}). The detail of the extraction of lattice constant and error is shown in the Supplemental Material\cite{SM}. The lattice constant of MLG directly beneath \WSe\ is determined to be $a_\mathrm{MLG}=2.4575\pm0.0007 \mathrm{\AA}$ from the same $\delta$-$\theta$ scans. Using these values, we deduce that on MLG/SiC, a 3$\times$3 unit cell of the compressed \WSe\ is perfectly commensurate with a 4$\times$4 graphene lattice, with an experimentally determined mismatch below 0.03 \%. This could explain why monolayer \WSe\ is compressed on MLG.

We now turn to the result of the first-principles calculation to examine the role of strain.
Figures \ref{fpc}(c-f) show how a compressive or tensile strain modifies the valence band structure of monolayer \WSe. In the respective calculations, the in-plane lattice constant was changed proportionally to include strain while keeping the unit cell volume constant.
Qualitatively, our calculations indicate that compressive strain reduces the value of $\Delta_\mathrm{SO}$ [Fig.~\ref{fpc}(f)], which is in accordance with previous first-principles results~\cite{Le2015}.
We find $\Delta_\mathrm{SO}$ = 452 (475) meV using Wien2k (ADF-BAND) for ``zero strain'', i.e., when we fix the lattice constants of monolayer identical to bulk. 
By introducing strain, inferred change in $\Delta_\mathrm{SO}$ is $+$ ($-$) 18 meV per 1 \% of tensile (compressive) strain as shown in Fig.~\ref{fpc}(f). This holds for moderately strained \WSe\ (as is the case in experiment) while the general dependence of $\Delta_\mathrm{SO}$ on strain is clearly nonlinear. 
Using the experimentally determined value of the lattice compression of WSe$_2$ on graphene ($-$0.19\,\%), the amount of change in $\Delta_\mathrm{SO}$ that could arise from compressed strain is $-$3.4 meV. This is smaller than the error in experimental $\Delta_\mathrm{SO}$ (469$\pm$8~meV), and much smaller than the difference of 44 meV between our epitaxial monolayer \WSe\ and the exfoliated one from a bulk crystal~\cite{Le2015}. We thus conclude that a strain effect cannot explain the discrepancy in $\Delta_\mathrm{SO}$.

A subtle issue in the approach we used to estimate strain is that we actually do not know the lattice constant of a freestanding monolayer \WSe. Namely, a monolayer \WSe\ even without substrate effects may not have an identical lattice constant as that of bulk counterpart. To examine this, we performed additional first-principles calculations for bulk and monolayer \WSe\ with structural optimization\cite{SM}. The theoretical results showed that the lattice constant of monolayer \WSe\ converges to almost identical value as that of bulk (expanded only by $\sim+$0.03\% \cite{SM}). This means that the experimentally observed compression ($-$0.19\%) could be attributed to a strain in monolayer \WSe\ as we have assumed.

Because strain in the epitaxial \WSe\ is excluded as an origin of discrepancy in $\Delta_\mathrm{SO}$, we point out alternative possibilities.
We first examined the possibility that the larger $\Delta_\mathrm{SO}$ observed in exfoliated \WSe\ came from a tensile strain in the flake when transferred to the substrate.
By using the second-order polynomial fit to the $\Delta_\mathrm{SO}$ vs. strain plot [Fig.\,\ref{fpc}(f)], we found the maximum gain in $\Delta_\mathrm{SO}$ predicted by the theory is $+40$\,meV for $+4.5$\,\% tensile strain.
This is close to the experimentally found difference ($\sim$48\,meV), which means that $+\sim$4-5\% of tensile strain in the exfoliated bulk is needed to reproduce the value observed in the exfoliated bulk by strain.
However, this is unlikely because the band dispersion of monolayer \WSe\ expected for such a large strain [see Fig.\,\ref{fpc}(e)], which is rather different from the pristine monolayer, is not observed in the ARPES of the exfoliated bulk \cite{Le2015}.
A very recent ARPES on exfoliated \WSe\ by Nguyen \textit{et al.} \cite{Nguyen2019} showed that (i) $\Delta_\mathrm{SO}$=0.485$\pm$0.01~eV for monolayer, much closer to the value observed in this study, and (ii) $\Delta_\mathrm{SO}$=0.501$\pm$0.01~eV for bilayer \WSe, which is close to that of earlier exfoliated monolayer result \cite{Le2015}. Thus, a more plausible origin for discrepancy may be that the larger $\Delta_\mathrm{SO}$ in the previous study was obtained due to some inclusion of bilayer \WSe\ in an exfoliated monolayer.

\subsection{Band alignment and charge transfer}\label{align}

\begin{figure}[!htb]
\includegraphics[width=0.95\linewidth,clip]{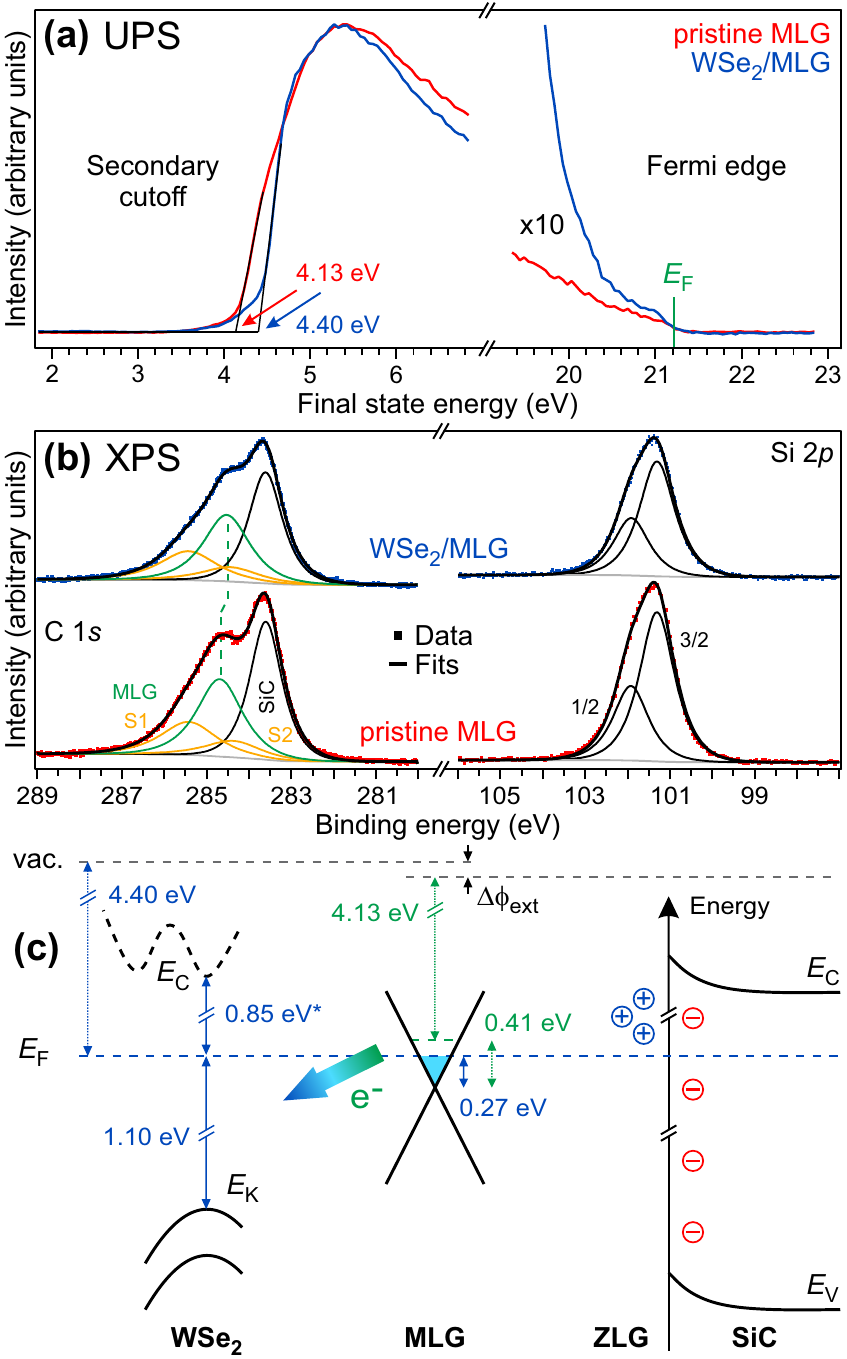}
\caption{(a) UPS spectra obtained from pristine MLG (red) and \WSe/MLG (blue). On the final-state-energy axis, the respective sample work function can directly be read off from the secondary cutoff (red and blue arrows). (b) XPS core level spectra of C 1$s$ and Si 2$p$. The shift of the MLG peak (green curve) to lower binding energies upon \WSe\ growth is consistent with the observed charge transfer from MLG onto \WSe. All other components are found unshifted, indicating that the band bending at the graphene/SiC interface is unperturbed by the \WSe\ growth. (c) Schematic band alignment of the \WSe/MLG heterostructure as obtained from photoelectron spectroscopy (not drawn to scale). The polarity contribution to the upward band bending at the SiC/ZLG interface (red circles) is partially compensated by donor states (blue circles). Electron transfer from graphene onto \WSe\ is indicated by the filled arrow. The Fermi energy before (after) the \WSe\ growth is shown by the green (blue) dashed lines. The additional contribution $\Delta\phi_\text{ext}$ to the work function change results from an upshift of the vacuum level due the change in surface termination from MLG to\WSe.
}
\label{ups}
\end{figure}

The sample work function $\phi$ can be measured using UPS. From the secondary cutoffs of the respective spectra as shown in  Fig.~\ref{ups}(a), we infer $\phi=4.13$ eV and 4.40 eV ($\pm 0.04$ eV) before and after the growth of \WSe, respectively. In combination with the ARPES results of Sec.~\ref{structstrain}, we derive the band alignment of the \WSe/MLG heterostructure as sketched in Fig.~\ref{ups}(c).
In quasi-free standing graphene, the bulk polarization of the SiC substrate induces an upward band bending, which would result in $p$-doping of the surface when terminated by a clean interface~\cite{Ristein2012}. Yet, with the presence of the buffer layer (ZLG) this is
overcompensated by donor states at the graphene/SiC interface, resulting in the $n$-type character of epitaxial MLG/SiC ~\cite{Mammadov2017} with its Dirac point residing 0.41 eV below \EF\ [see Fig.~\ref{arpes}(e)].
As discussed in Sec.~\ref{structstrain}, the Dirac point shifts up by 0.14 eV to 0.27 eV below \EF\ upon \WSe\ growth [see Fig.~\ref{arpes}(f)]. To our knowledge, such a shift of the graphene $\pi$-bands upon TMD growth on top was not reported previously. There are two possible mechanisms to explain this observation. First, electron transfer from graphene to \WSe\ could shift the graphene bands upwards. Second, if the donor states at the graphene/SiC interface are partially compensated during the TMD growth (e.g.\ via chemical reaction with the Se vapor), the amount of $n$-type doping of graphene could change. In the latter case, modified donor states should influence the band bending at the graphene/SiC interface, which can be detected via a shift of the SiC core levels. The fitted XPS core level spectra of C $1s$ and Si $2p$ are shown in Fig.~\ref{ups}(b). The C $1s$ fits consist of four components representing bulk SiC, MLG and the carbon buffer layer with its partial bonding to SiC (S1 and S2)~\cite{Riedl2010}. The Si $2p$ spectra can reasonably well be fitted by one spin-orbit split doublet ($j=3/2$ and $1/2$ with an area ratio of 2:1). We find that the SiC peaks are unshifted in energy after the growth of \WSe, indicating that the band bending at the interface remains unchanged. From this, we can exclude that the reduced $n$-type doping of graphene results from a modification of the interfacial donor states during TMD growth. Upon \WSe\ growth the C $1s$ MLG component shifts by $0.16\pm0.02$ eV to lower binding energies while S1 and S2 retain their positions. This core level shift of MLG is quantitatively in line with the upshift of the Dirac point observed in ARPES and further supports the scenario of electron transfer from MLG to \WSe. The work function increases by the charge transfer, and we ascribe the remaining increase of $0.27 - 0.14 = 0.13$ eV to an extrinsic upshift $\Delta\phi_\text{ext}$ of the vacuum level due to the change in surface termination from MLG to \WSe [Fig.~\ref{ups}(c)].
The valence band maximum $E_\mathrm{K}$ of \WSe\ is found $\sim 1.1$ eV below \EF\ in our ARPES measurements [see Fig.~\ref{arpes}(i)], which matches very well with the results obtained from MBE-grown epitaxial \WSe\ on bilayer graphene~\cite{Zhang2016}.
By assuming a band gap of $1.95$ eV as previously determined by STS~\cite{Zhang2016}, we estimate that the conduction band minimum $E_\mathrm{C}$ is located $\sim 0.85$ eV above \EF. As such, the Fermi level in \WSe\ resides closer to the conduction band minimum than to the valence band maximum. We finally note that a finite density of in-gap states can be expected in our epitaxial \WSe\ films, stabilizing the position of \EF\ inside the band gap after the electron transfer from graphene.

\subsection{Morphology of monolayer \WSe}\label{morph}

\begin{figure}[!t]
\includegraphics[width=0.9\linewidth,clip]{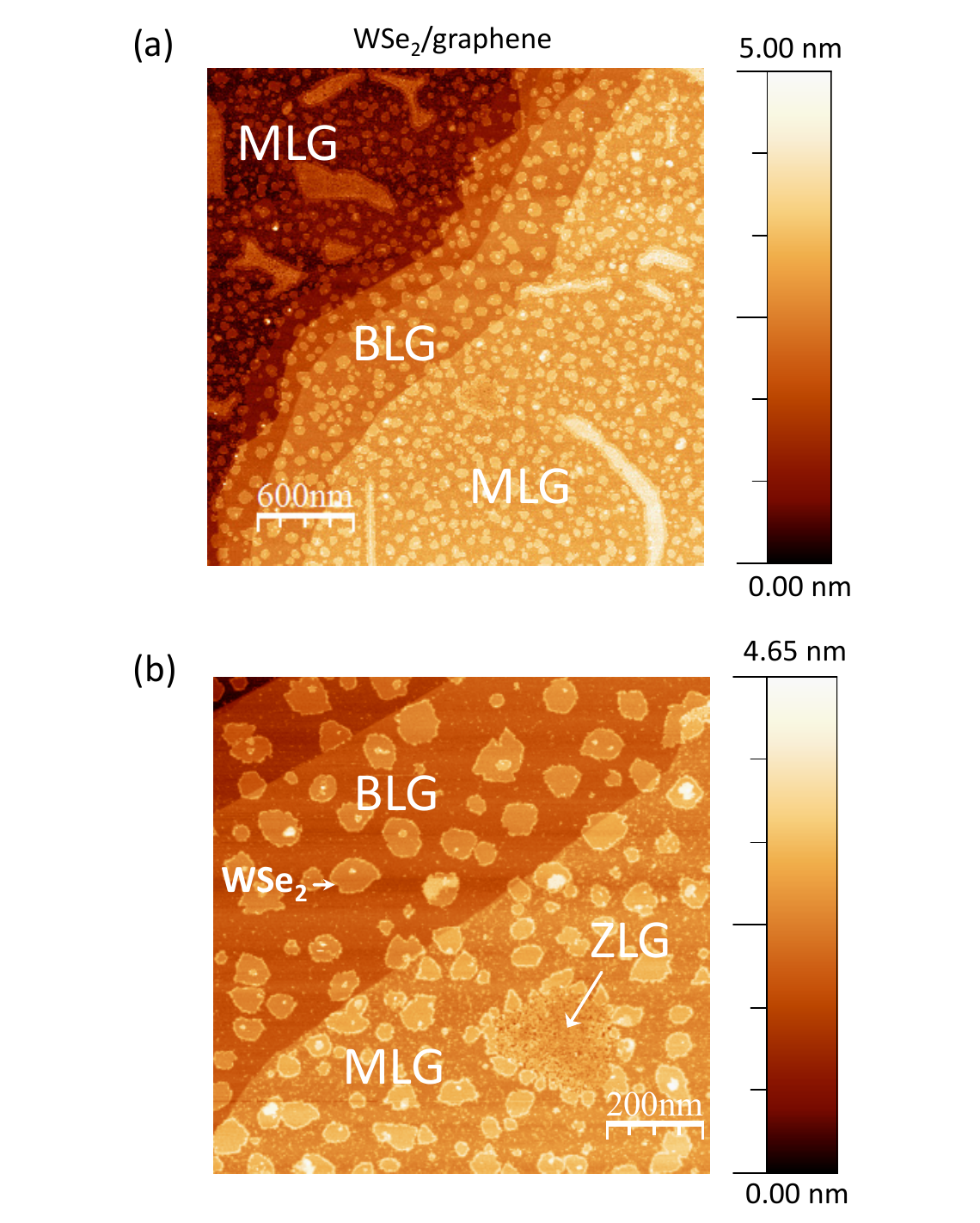}
\caption{Topographic AFM images of epitaxial \WSe\ on MLG/SiC. Lateral dimensions are (a) $3\times3$ $\mathrm{\mu}$m and (b) $1\times1$ $\mathrm{\mu}$m.}
\label{afm}
\end{figure}

The morphology of monolayer \WSe\ was measured by AFM.
The unique feature of the $h$-PLD grown films was a high spatial uniformity with relatively small island sizes.
Larger scale AFM images show a high density of nucleation sites distributed uniformly over the surface [Fig.\,\ref{afm}(a)].
The epitaxial graphene substrate had a minor inhomogeneity on the surface coming from the fabrication process, resulting in only small areas of bilayer graphene (BLG) close to the step edge, and other areas of exposed buffer layer (ZLG, which lacks a Dirac linear dispersion) within the flat MLG terrace region.
Notably, \WSe\ islands were indiscernible on ZLG by AFM [Fig.\,\ref{afm}(b)].
On the other hand, islands on BLG tended to be larger (frequently approaching $\sim$100 nm) than on MLG [Fig.\,\ref{afm}(b)].
The different \WSe\ island sizes throughout the epitaxial graphene substrate are likely related to the distinct chemical nature and morphology of BLG, MLG and ZLG.
During the TMD growth process, the migration of species could be severely limited on ZLG due to their covalent bonding to SiC and the resultant buckled surface~\cite{sforzini}, in contrast to the weak interaction on a complete vdW layers (MLG and BLG).
For the latter, BLG regions may have even smoother surface than that of MLG regions due to the remoteness to the covalent bonds.
Thus, our result clearly highlights the advantage of a chemically inert and smooth vdW surface in obtaining larger \WSe\ domains during the epitaxial growth.

\section{Summary and conclusions\label{sum}}
A spin splitting of $\Delta_\mathrm{SO}=0.469\pm0.008$~eV is found for the topmost \WSe\ valence band at $\overline{\mathrm{K}}$. The in-plane lattice constant of \WSe\ was determined by grazing incidence X-ray diffraction, revealing a small compression ($-0.19$ \%) of the epitaxial monolayer \WSe\ film with respect to its bulk counterpart. Supplementing these data with first-principles calculations, we conclude that potential strain effects on $\Delta_\mathrm{SO}$ are negligible in our \WSe\ film. Furthermore, the overall band alignment between \WSe\ and graphene was clarified. The electron transfer from graphene to \WSe\ becomes apparent from an upshift of the Dirac point of graphene with respect to the Fermi level after the growth of the TMD monolayer.
The varying \WSe\ island sizes on substrate areas covered by graphene layers of different thicknesses suggest the importance of atomically smooth, weakly interacting van der Waals surfaces for monolayer TMD epitaxy.
Our results provide high-quality data on both electronic and structural properties of monolayer \WSe\ and shed light on potential substrate influences in bottom-up TMD growth.

\begin{acknowledgments}
We are grateful to D. Huang and D. Weber for discussions and critical reading of the manuscript. We thank S. Prill-Drimmer, K. Pflaum, M. Dueller, and F. Adams for technical support.
We acknowledge Diamond Light Source for time on Beamline I07 under Proposal SI18887 and Calipso program for the financial support. This work was supported by the Alexander von Humboldt-Foundation.

\end{acknowledgments}

\end{document}

\title{%
Supplemental Material for\\
``Spin splitting and strain in epitaxial monolayer WSe$_2$ on graphene''
}
\author{H. Nakamura}
\email{hnakamur@uark.edu; present address: Department of Physics, University of Arkansas, AR 72701, USA}
\affiliation{Max Planck Institute for Solid State Research, 70569 Stuttgart, Germany}
\author{A. Mohammed}
\affiliation{Max Planck Institute for Solid State Research, 70569 Stuttgart, Germany}
\author{P. Rosenzweig}
\affiliation{Max Planck Institute for Solid State Research, 70569 Stuttgart, Germany}
\author{K. Matsuda}
\affiliation{Max Planck Institute for Solid State Research, 70569 Stuttgart, Germany}
\affiliation{Nagoya University, Nagoya 464-8603, Japan}
\author{K. Nowakowski}
\affiliation{Max Planck Institute for Solid State Research, 70569 Stuttgart, Germany}
\affiliation{University of Twente, 7522 NB Enschede, Netherlands}
\author{K. K{\"u}ster}
\affiliation{Max Planck Institute for Solid State Research, 70569 Stuttgart, Germany}
\author{P. Wochner}
\affiliation{Max Planck Institute for Solid State Research, 70569 Stuttgart, Germany}
\author{S. Ibrahimkutty}
\affiliation{Max Planck Institute for Solid State Research, 70569 Stuttgart, Germany}
\author{U. Wedig}
\affiliation{Max Planck Institute for Solid State Research, 70569 Stuttgart, Germany}
\author{H. Hussain}
\affiliation{Diamond Light Source Ltd, Didcot, Oxforshire OX11 0DE, United Kingdom}
\author{J. Rawle}
\affiliation{Diamond Light Source Ltd, Didcot, Oxforshire OX11 0DE, United Kingdom}
\author{C. Nicklin}
\affiliation{Diamond Light Source Ltd, Didcot, Oxforshire OX11 0DE, United Kingdom}
\author{B. Stuhlhofer}
\affiliation{Max Planck Institute for Solid State Research, 70569 Stuttgart, Germany}
\author{G. Cristiani}
\affiliation{Max Planck Institute for Solid State Research, 70569 Stuttgart, Germany}
\author{G. Logvenov}
\affiliation{Max Planck Institute for Solid State Research, 70569 Stuttgart, Germany}
\author{H. Takagi}
\affiliation{Max Planck Institute for Solid State Research, 70569 Stuttgart, Germany}
\affiliation{Department of Physics, University of Tokyo, 113-0033 Tokyo, Japan}
\affiliation{Institute for Functional Matter and Quantum Technologies, University of Stuttgart, 70569 Stuttgart, Germany}
\author{U. Starke}
\affiliation{Max Planck Institute for Solid State Research, 70569 Stuttgart, Germany}

\maketitle

\begin{widetext}

\subsection{Spin splitting determined by ARPES}

We show detailed analysis of an error related to the fitting of energy distribution curve (EDC) in ARPES. In the main text [Fig. 2(a)], we presented pseudo-Voigt function fits to the experimental EDC. A linear background was subtracted from EDC prior to the fitting. Here, we present additional fitting results where we changed the energy range used for fitting to check the influence of linear background subtraction. The result is shown in Fig.~S1. The two fits produced $\Delta_\mathrm{SO}$ = 0.468~eV and 0.474~eV, differing by $\sim$6meV.

Next, we performed EDC fits using Gaussian functions. The result is shown in Fig.~S2. The two fits produced $\Delta_\mathrm{SO}$ = 0.459~eV and 0.476~eV, differing by $\sim$17~meV. The $\Delta_\mathrm{SO}$ error from Gaussian fits is more prone to the change in an assumed background, probably because the fits to the peak top is less satisfactory than the pseudo-Voigt fits.

Finally, by averaging these four fitting results and by using standard deviation for error estimate, we obtain $\Delta_\mathrm{SO}=469\pm$8~meV.

\begin{figure*}[!h]
\includegraphics[width=13cm,clip]{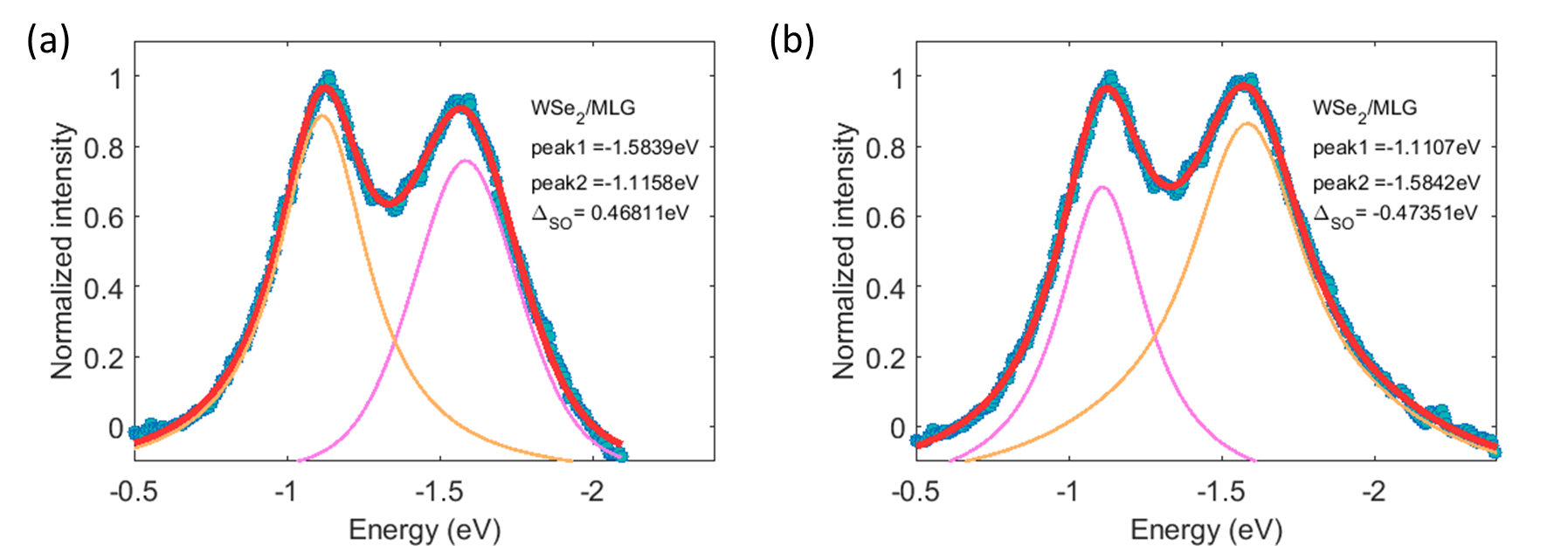}
\caption{%
EDC fits at the K point of \WSe\ using pseudo-Voigt functions for (a) a small energy range (-0.5~eV to -2.1~eV) and (b) a large energy range (-0.5~eV to -2.5~eV). The blue filled circles show experimental data, and the red solid lines shows fits assuming two peaks (shown in solid orange and magenta lines).
}
\label{fig:A}
\end{figure*}

\begin{figure*}[!h]
\includegraphics[width=13cm,clip]{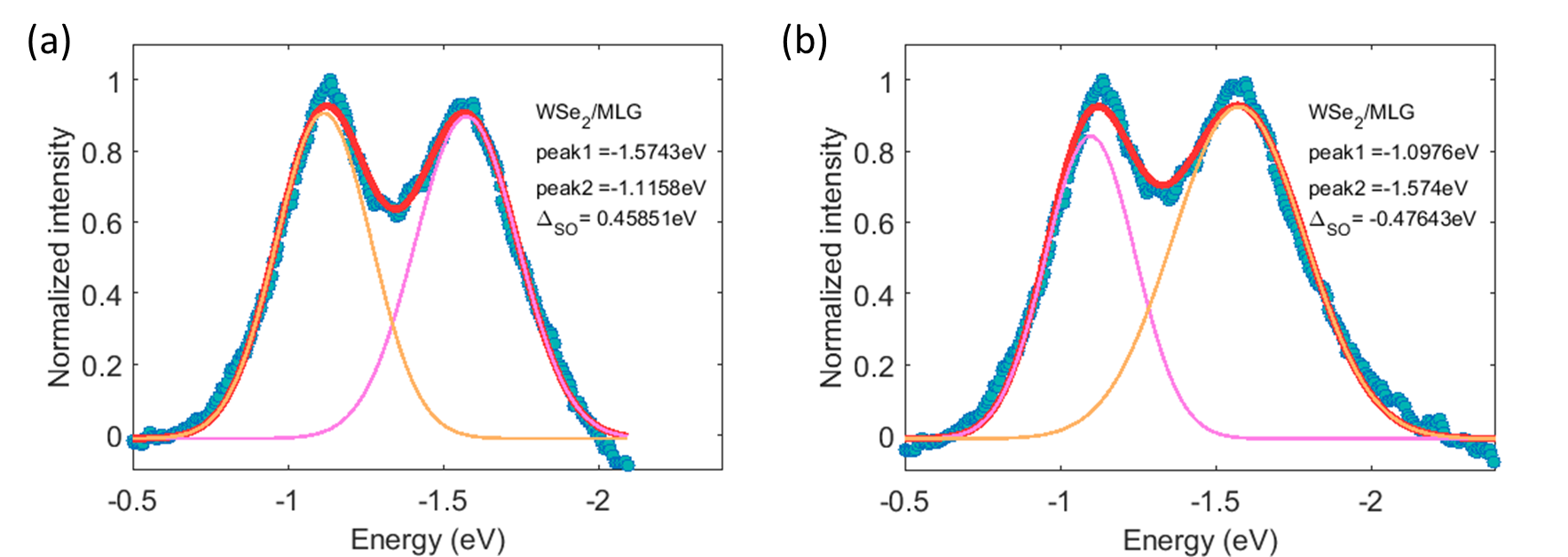}
\caption{%
EDC fits at the K point of \WSe\ using Gaussian functions for (a) a small energy range (-0.5~eV to -2.1~eV) and (b) a large energy range (-0.5~eV to -2.5~eV). The blue filled circles show experimental data, and the red solid lines shows fits assuming two peaks (shown in solid orange and magenta lines).
}
\label{fig:B}
\end{figure*}

\pagebreak

\subsection{Lattice constant derivation from GIXRD}

We describe detail of XRD analysis used to derive lattice constants.
First, we obtained precise diffraction angles of monolayer \WSe\ by fitting all peak positions by Gaussian functions (Fig. S3). Next, the plane spacing was derived for each Miller index via the Bragg’s law ($2d\sin\theta= \lambda$, where $\lambda = 1.0332\mathrm{\AA}$). Finally, the in-plane lattice constant was calculated by assuming the hexagonal lattice using the following formula:
\begin{equation}
\frac{1}{d^2} = \frac{4}{3}\left( \frac{h^2+hk+k^2}{a^2}+\frac{l^2}{c^2} \right).
\end{equation}
where $h$, $k$, $l$ are the Miller indices, $a$ and $c$ are the in-plane and out-or-plane lattice constant, respectively. Note that in our GIXRD, $l$=0. The lattice constant for each diffraction angle obtained this way is shown in Fig.~S4.

\begin{figure*}[!h]
\includegraphics[width=17cm,clip]{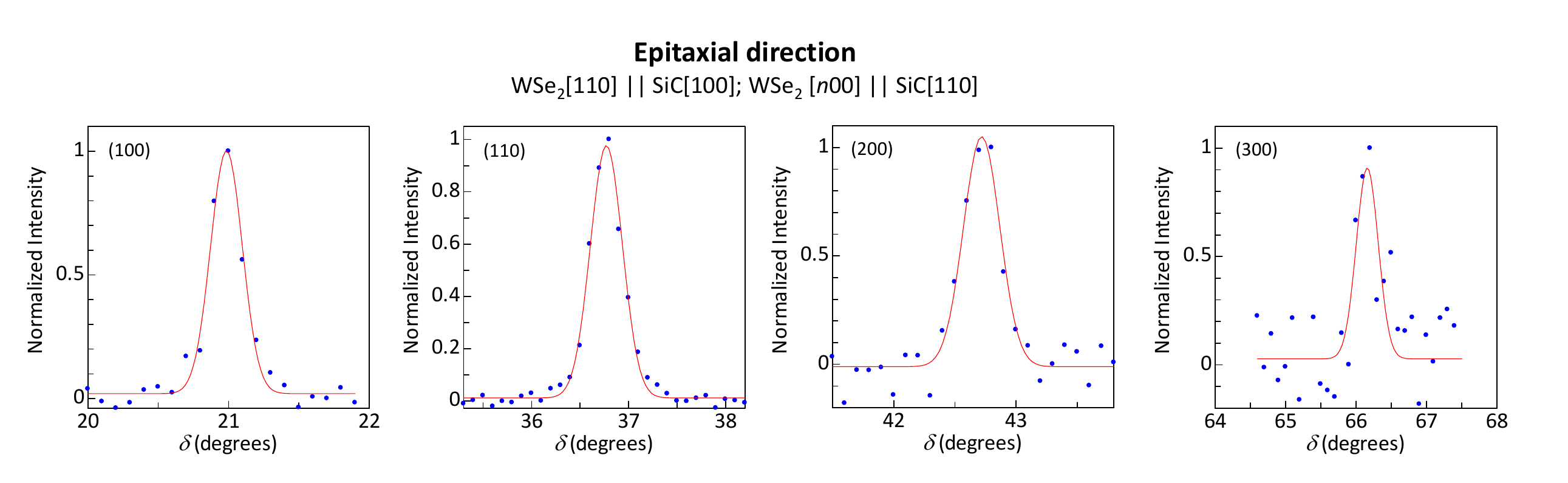}
\caption{%
Monolayer \WSe\ diffraction peaks from GIXRD $\delta-\theta$ scan. Diffraction angles are determined by a scan along epitaxial direction for each Miller index with respect to the MLG/SiC substrate (i.e. Q$\parallel$SiC[110] for \WSe(110), Q$\parallel$SiC[100] for \WSe($n$00), where $n$=1, 2, and 3). The experimental data points (blue circles) were fit by the Gaussian curves (red lines) to obtain peak position.
}
\label{fig:C}
\end{figure*}

\begin{figure*}[!h]
\includegraphics[width=7cm,clip]{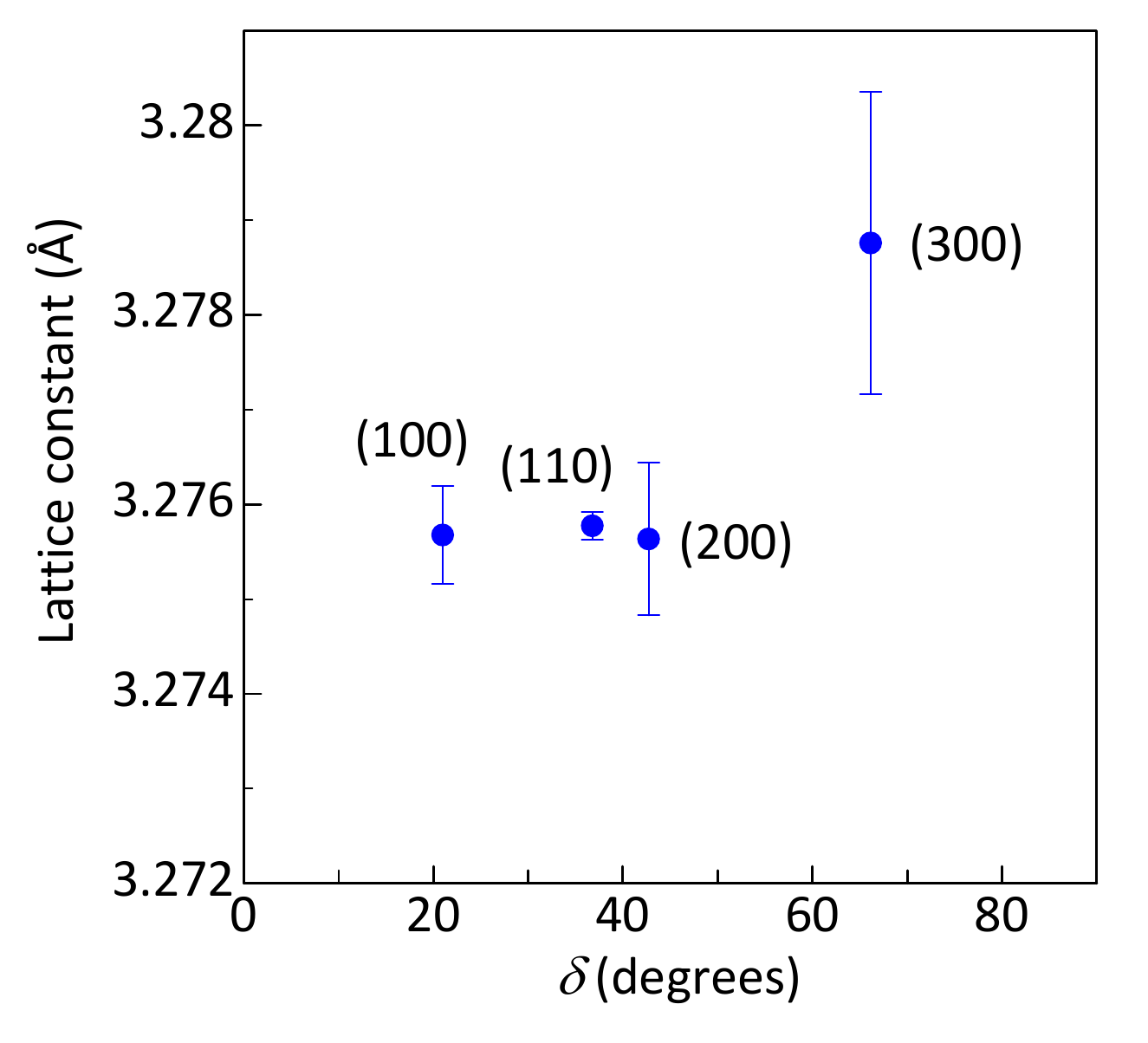}
\caption{%
The in-plane lattice constant of monolayer \WSe\ derived from each peak position in GIXRD $\delta-\theta$ scan.
}
\label{fig:D}
\end{figure*}

An error associated with determining the peak position is examined by adding a synthesized Poisson noise whose characteristic is matching that of a real data \cite{Wilson}. To do so, we first checked that a real background data indeed obeyed Poissonian statistics, by analyzing a data nearby each peak (excluding the region with a peak). Second, we added a random Poisson noise generated by a computer code to each diffraction peak, and performed a Gaussian fit to extract a peak position from the synthetic data. The procedure is repeated ten times for each Miller index, which is equivalent of doing ten experiments in the same condition. By this procedure, we can estimate how prone the extracted peak position (and thus, the derived lattice constant) is to the count statistics. Finally, an error bar for the lattice constant of each Miller index is calculated by standard deviation of results from the ten data set. The error obtained this way is reflected in the error bar of Fig.~S4.

A precision in determining the lattice constant improves at higher diffraction angles when the main source of error comes from angular error \cite{Cullity}. However, in the present case, (i) the (300) peak has the worst signal to noise ratio (see Fig. S3) and thus the worst precision in lattice constant (Fig.~S4), and (ii) no clear trend in lattice constant as a function of diffraction angle is found for three diffraction peaks (100),(110), (200), for which we had good statistics, indicating that an angular error may not be the main cause of error.

To obtain the value of lattice constant of \WSe, we took an average of those obtained for three diffraction peaks: (100), (110), and (200) in Fig.~S4. The (300) peak was omitted due to poor precision. As for error, we used that of (200), which was the worst among the three data used in averaging. The result is 3.2757$\pm$0.0008~$\mathrm{\AA}$. The magnitude of error corresponds to 0.03\% in lattice constant.

Next, we discuss the accuracy of lattice constant determined by GIXRD. The diffraction from SiC substrate recorded in the same $\delta-\theta$ scan is used for this purpose. By adopting the same approach as for \WSe, we extract the lattice constant of SiC from each peak position in the in-plane $\delta-\theta$ scan. The result is shown in Fig. S5.

\begin{figure*}[!h]
\includegraphics[width=7cm,clip]{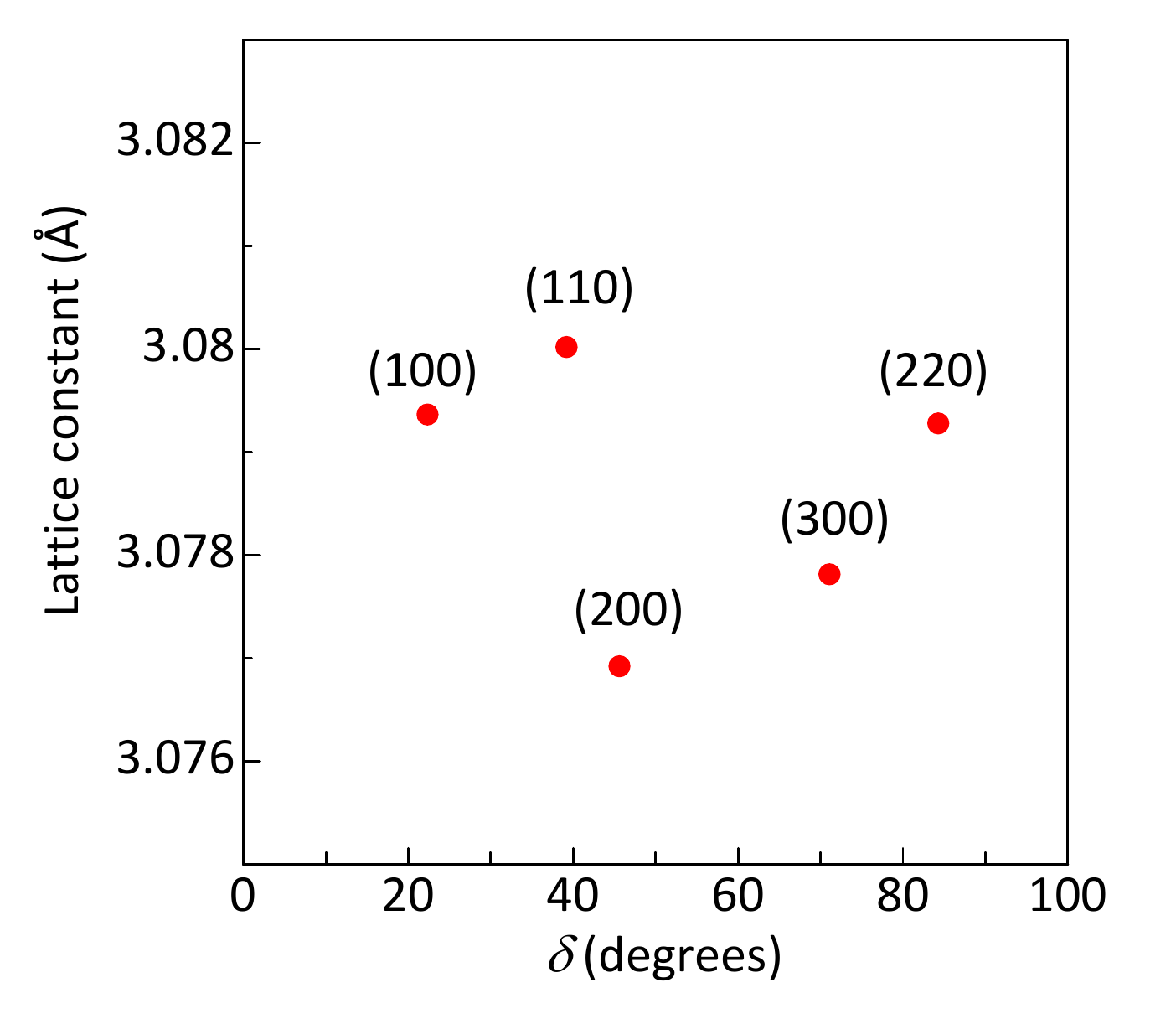}
\caption{%
The in-plane lattice constant of SiC derived from each peak position in GIXRD $\delta-\theta$ scan.
}
\label{fig:E}
\end{figure*}

We then averaged the five lattice constant values to derive the lattice constant of SiC, and used standard deviation of the five data points for an error. We obtain $a_\mathrm{SiC}=3.0787\pm0.0013 \mathrm{\AA}$, corresponding to $\pm$0.04\% precision. A slightly larger error compared to \WSe\ originates not from a count statistics, but by having only a few data points in the $\delta-\theta$ scans available for Gaussian fits due to sharpness of SiC peaks. Among several literature for SiC, the work by Hass \textit{et al}. \cite{Hass} reports the lattice constant of an un-doped, synthetic 4H-SiC, similar to what we used, to be $a_\mathrm{SiC}=3.0791\pm0.0006 \mathrm{\AA}$. The two values match within the error of our measurement. Thus, the \textit{accuracy} of the lattice constant from our GIXRD is estimated to be approximately $\pm$0.04\%.

To estimate a ``strain'' with respect to bulk \WSe, we adopted the lattice parameter from Schutte \textit{et al}.\cite{Schutte} as a bulk reference. The reported lattice constant of 2H-\WSe\ is $a=3.282\pm0.001\mathrm{\AA}$ \cite{Schutte}, corresponding to a precision of 0.03\%.
Using the lattice constant for \WSe\ ($a_\mathrm{WSe_2}=3.2757\pm0.0008\mathrm{\AA}$), the “strain” of the monolayer \WSe\ with respect to the bulk reference is $\left(\frac{3.2757}{3.282}- 1\right) \times100 = -0.19 \%$. We note that errors associated with this estimation [our GIXRD (0.03\%); bulk XRD from Ref. \cite{Schutte} (0.03\%)] are both appreciably smaller than the degree of lattice compression found here.

\subsection{First principles calculations with relaxed lattice}

We present structure optimization results of bulk and monolayer \WSe\ using the first-principles calculations as implemented in ADF-BAND \cite{AMS,teVelde,BAND,Kadantsev} in Table I. We identify small lattice expansion (+0.03\%) from bulk to monolayer when compared within a model that includes spin-orbit coupling (SOC). This means that within the theory, free-standing monolayer should have almost identical lattice constant as that of a bulk counterpart. The spin splitting in the valence band at $\overline{\mathrm{K}}$ was 0.475 eV (monolayer w/SOC, bulk structure) and 0.471 eV (monolayer w/SOC, optimized structure), respectively.

\begin{table}[!h]
\caption{\label{tab:table1} First-principles structure optimization results for \WSe. The lattice compression/expansion in percentage (red) is calculated with respect to the experimental lattice constant for bulk\cite{Schutte}.}
\begin{ruledtabular}
\begin{tabular}{llllll}
--  & $a (\mathrm{\AA}$) \textcolor{red}{(\%)}& $c (\mathrm{\AA}$) \textcolor{red}{(\%)}& Band gap\footnote{Band gap, indirect (i) or direct (d)} (eV) & $r_\mathrm{W-Se}$\,($\mathrm{\AA}$) & $r_\mathrm{Se-Se}$\footnote{distance of Se-Se perpendicular to \WSe\ layer (a ‘layer thickness’)} ($\mathrm{\AA}$)
\\
\hline
Bulk rel. w/SOC, exp. structure\footnote{Schutte et al. [1]}& 3.282 & 12.96 & 0.898(i) \footnote{Scalar relativistic band gap for experimental structure: 0.967 eV (i).} & 2.526 &3.341
\\
Bulk scalar rel., opt. structure & 3.278 \textcolor{red}{(-0.12)} & 13.276 \textcolor{red}{(+2.4)}& 1.073 (i) & 2.533 &3.368
\\
Bulk rel. w/SOC, opt. structure & 3.277 \textcolor{red}{(-0.15)} & 13.281 \textcolor{red}{(+2.5)}& 0.977 (i) & 2.533 &3.369
\\
2ML scalar rel., opt. structure & 3.278 \textcolor{red}{(-0.12)} & --  & 1.351 (i) & 2.532 &3.366
\\
ML scalar rel., opt. structure & 3.275 \textcolor{red}{(-0.21)} & --  & 1.655 (i) & 2.532 &3.368
\\
ML w/SOC, opt. structure & 3.278  \textcolor{red}{(-0.12)} & -- & 1.323 (i) & 2.533 & 3.366
\\
\end{tabular}
\end{ruledtabular}
\end{table}

\end{widetext}